\documentclass[aps,pre,twocolumn,groupedaddress,showpacs,floatfix]{revtex4}
\usepackage{amsmath}
\usepackage{amssymb}
\usepackage{graphicx}

\begin{document}

\title{Voronoi and Fractal Complex Networks and Their Characterization}
\author{Luciano da Fontoura Costa}
\email[]{luciano@if.sc.usp.br}
\affiliation{Institute of Physics of S\~ao Carlos. University of S\~ ao Paulo, S\~{a}o Carlos, SP, Caixa Postal 369, 13560-970, Brasil}

\date{\today}

\begin{abstract}   
Real complex networks are often characterized by spatial constraints
such as the relative position and adjacency of nodes.  The present
work describes how Voronoi tessellations of the space where the
network is embedded provide not only a natural means for relating such
networks with metric spaces, but also a natural means for obtaining
fractal complex networks.  A series of comprehensive measurements
closely related to spatial aspects of these networks is proposed,
which includes the effective length, adjacency, as well as the fractal
dimension of the network in terms of the spatial scales defined by
successive shortest paths starting from a specific node.  The
potential of such features is illustrated with respect to the random,
small-world, scale-free and fractal network models.
\end{abstract}


\maketitle

Several natural objects and phenomena can be modeled in terms of 
networks of basic components intensively exchanging information between
themselves.  Recent findings about the behavior of real networks,
noticeably the small-world and scale-free
models\cite{Albert_Barab:2002}, motivated a renewal of interest in
mathematical and statistical investigations of the properties of such
complex structures.  These studies often consist of estimating
fundamental properties of the network, such as the degree
distribution, clustering coefficient, and average length, for several
network parameter configurations.  For instance, the degree
distribution of networks where the probability of new connections is
proportional to the node degree has been verified to follow a power
law \cite{Albert_Barab:2002}, indicating scale invariance, hence the
name \emph{scale-free networks}.  While such investigations have led
to a series of interesting findings, relatively little attention, with
a few exceptions \cite{Waxman:1988,Manna:2003}, has been given to the
role of spatial relationships between the network nodes as constrains
to the network topology.  Indeed, connections in real networks are
often inherently related to the spatial positions of the constituent
nodes as well as to their spatial adjacencies.  Real networks falling
into this category include but are not limited to telephone and
electric power distribution, transportation systems, computer
networks, spatio-temporal gene activation in animal development, as
well as the neuronal networks in the mammals retina and cortex.

The following three main kinds of spatial requirements are considered
here: (i) connect adjacent or near nodes, (ii) minimize the access
(according to some cost such as time or distance) to more distant
regions of the net and (iii) minimize the number and/or total length
of edges.  As such items are often incompatible -- e.g. minimizing the
access between any pair of nodes implies a large number of edges,
networks are often characterized by a compromise between such
requirements.  One of the purposes of the current work is to
investigate how the topologies of complex networks are related to
spatial constraints.  In other words, as the topology of real networks
is often optimized to suit specific tasks, it becomes particularly
important to obtain measurements directly expressing such
spatial/function relationships.

Traditional random, small-world and scale free networks, as well as
fractal networks, are considered in this paper.  The network nodes are
assumed to be spatially distributed along the two-dimensional region
$\Omega$ according to the Poisson distribution with parameter $p$.
The domain $\Omega \subset R^2$ where the network is embedded is
assumed to be the square $(0 \leq x \leq L, 0 \leq y \leq L)$.  The
Euclidean distance between any two nodes $i$ and $j$ is represented as
$d_{i, j}$.  Each point $(x, y)$ of $\Omega$ is associated to the
nearest network node, which defines a Voronoi partition of $\Omega$
\cite{ahuja:1983}, and consequently a \emph{Voronoi network}.  The
spatial position of each network node $i$, $i=1,2, \ldots, N$, is
expressed as $P_i=(x_i, y_i)$, and the area of the Voronoi cell
respectively associated to this point is henceforth represented as
$A_i$.  Given the initial nodes, any type of network can be
constructed by adding connections between them.  The degree of a
specific node $i$ is $k_i$, and the cost (or length) of the edge
between nodes $i$ and $j$, hence $\lambda_{i, j}$, is given by the
Euclidean distance between those nodes.  As the networks considered
here are not oriented, we have that $\lambda_{i, j}=\lambda_{j,
i}$. Infinite length is assigned to inexistent edges and not taken
into account for the statistics.  The shortest path between two nodes
$i$ and $j$, expressed as $\varepsilon_{i, j}$, is given by the sum of
the edge lengths composing that shortest path.

\begin{figure}
 \includegraphics[scale=.35,angle=-90]{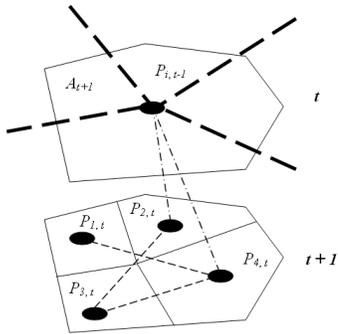} 
 \caption{A Voronoi cell defined by node $P_{i,t-1}$ at hierarchy $t$
 and a daughter network and respective Voronoi partition at the
 subsequent hierarchical level.~\label{fig:fracnet}}
\end{figure}

The concept of hierarchy provides a powerful organizational framework
to model several real networks
\cite{Dorogovtsev:2002,Barab_Oltvai:2003,Ravasz_Barab:2002}.  Such a
structuring can be immediately extended to Voronoi networks.  Indeed,
the tessellation of the network space $\Omega$ provides a natural and
generic basis for defining a broad class of spatially constrained
hierarchical and fractal networks.  The basic construction principle
is to define a new network inside each of the current Voronoi cells,
as illustrated in Figure~\ref{fig:fracnet}.  Each added hierarchy is
identified by successive integer indices $t= 0, 1, \ldots, Q$, and it
is henceforth assumed that the nodes can only make connections inside
their respective Voronoi cell and with the parent node, to which they
are connected according to some statistical role, e.g. obeying uniform
distribution with fixed parameter $\gamma_m$, as adopted henceforth.
At least one random connection is guaranteed between the parent cell
and one of the nodes of the daughter cell.  In this way, the overall
network can be understood as a series of spatially congruent networks
connected along the hierarchical levels. The parameter $\gamma_m$
controls the level of interaction between distinct hierarchies.  In
case $\gamma_m$ is low, the subnetworks behave in an almost
independent fashion.  A broad variety of connecting models can be
adopted inside each Voronoi cell, including random, small-world and
scale-free approaches.  Although hybrid networks can be obtained by
using different connecting schemes at each hierarchical level or cell,
the fractal networks in this article are all homogeneous and random.

Different fractal networks can be obtained by varying the nodes
distribution inside each cell and the way in which the connections
between the nodes are established. It is henceforth assumed that each
Voronoi cell along the hierarchies have the same average number of
nodes $M$ throughout, so that the average total number of nodes at
level $t$ is $M^{t+1}$.  Since $M$ is maintained throughout the
network, cells at subsequent hierarchies become progressively denser,
implying smaller spatial scale and increased level of details.  The
connections at any hierarchical level are uniformly established with
the same probability $\gamma$, and it is reasonable to make $\gamma =
\gamma_m$.  The network starts from hierarchy $t=0$ with the $M$ nodes
being connected with probability $\gamma_m$ to the initial master node
$P_0$.  As the distribution of points inside each cell at level $t$
obeys a Poisson distribution with parameter $p_{t}$, we have $M =
p_{t} \left< A_t \right>$ and $\left< A_t \right>=1/ (p_{t-1})$
\cite{ahuja:1983}, so that $p_t=p_{t-1}M=p_0 M^{t}$.  Observe that
$A_0=L^2$. The tessellations extend up to hierarchy $t=Q$ such as
$p_Q<1$.  Therefore $Q<-log(p_0-M)$.  The average total number of
nodes at level $t$ is $N_t=p_t A_0$, so that the average total amount
of nodes in the network is $N = 1+\sum_{t=0}^{Q} p_t A_0$.  The
average node degree at any cell in hierarchy $t$ is $\left< k_t
\right> \cong \gamma M$. The characteristics of the network are
therefore completely determined by the choice of $L$, $\gamma$, $M$
and $p_0$.

The self-similar nature of such fractal random networks, allied to the
fact that several of the properties inside each cell is independent of
those in other cells, allow convenient estimation of many network
features.  In case the distribution of a measurement $w$ inside any
cell at any cell in level $t$ can be expressed in terms of $t$, i.e.
$P_t(w)$, it is possible to estimate its distribution from level $0$
to $b<Q$ by using Equation~\ref{eq:distr}.

\begin{equation}
  P(w) = \frac{1}{\sum_{t=0}^{b} c_t} \sum_{t=0}^{b} c_t P_t(w)   \label{eq:distr} 
\end{equation}

where $\left< c_t \right> = A_0 p_0 M^{t-1}$ is the average number of
Voronoi cells at level $t$.  In case $P_t(w)$ is constant with respect
to $t$ and not too dispersed, as can be the case with $\left< n_t
\right>$ and $\left< k_t \right>$ under certain circumstances
\cite{Albert_Barab:2002}, the resulting distribution will lead to the
same distribution $P_t(w)$, now more intensely sampled.  For instance,
in the case of fractal random networks, we have that the degree
distribution for $N$ nodes will approach that of a random network with
$M$ nodes.  In other words, the degree distribution of the fractal
network does not follow a power law and is smaller than for a random
network with the same overall number of nodes.  The characteristic
spatial scale at level $t$, hence $r_t$, can be defined as the radius
of the equivalent circle with area $A_t$.  Consequently, $r_t=(\pi
p_0)^{-0.5} M^{0.5(1-t)}$. Any network property directly proportional
to $r_t$, as is the case with the edge length, shortest path and
nearest neighbor distances, will therefore follow a power law along
$t$ with exponent $-0.5log(M)$.  Such a property will also be
characterized by unit fractal dimension with respect to
$s_t=log(r_t)$.

The following measurements are suggested and used in the present work
to characterize the spatial properties of complex networks in a more
complete fashion:

\emph{Edge distance:} Although the average length, meaning the average
shortest path between any two network nodes, has been traditionally
used in order to characterize complex networks
\cite{Albert_Barab:2002}, the distribution of the Euclidean distances
between two directly connected nodes $i$ and $j$ can also be provide
useful information about the spatial characteristics of the network.
The edge distance is henceforth expressed as $\lambda_{i,j}$, with
average $\left< \lambda \right>$.

\emph{Effective distance:} This feature applies to each pair of nodes
$i$ and $j$, defined as the sum $S$ of the Euclidean distances of the
edges composing the shortest path between those nodes, in case it
exists, and the Euclidean distance between them,
i.e. $E_{i,j}=S/\lambda_{i,j}$.  Its maximum value of 1 is achieved
whenever the shortest path equals the Euclidean distance, implying
that all edges are parallel to the edge between $i$ and $j$.  Large
values of $E_{i,j}$ indicate network adherence to criterion (ii).

\emph{Adjacency:} This measurement, represented as $\alpha_i$, is
defined for each node $i$ as the number of direct connections it
establishes with its $n_{adj}$ spatially adjacent neighbors divided by
$n_{adj}$.  Thus $0 \leq \alpha_i \leq 1$.  Although slightly similar
to the clustering coefficient (e.g. \cite{Albert_Barab:2002}), this
feature differs in the sense that it takes the spatial adjacency
explicitly into account. The average of this feature for all network
nodes, i.e. $\left< \alpha \right>$, can be used to quantify criterion
(i).

\emph{Network fractal dimension:} While the average length and network
diameter \cite{Albert_Barab:2002} have been traditionally applied to
quantify the edges length distribution along the network, we present
an additional measurement, related to the multiscale spatial dimension
\cite{PRE_Marconi:2003}, which is capable of expressing the
self-similarity/fractality of the network as the shortest distance
nodes are found while the network is inundated (or visited) starting
from any specific node $i$. The network fractal dimension, therefore,
is a function of the spatial scale $s=log($Euclidean distance$)$,
defined for each node $i$ and represented as $f_i(s)$.  First, the
sorted sequence ${p_{i, 1}, \ldots, p_{i, j}, \ldots, p_{i, N}}$ of
shortest paths (considering the Euclidean distance and not the number
of nodes) between $i$ and every other node $j$ in the network is
obtained by using traditional methods.  The area distribution of node
$i$ can now be defined as $h_i(s) = \sum_{j=0}^{N} A_i
\delta(s-log(p_{i,j}))$, where $\delta$ is Dirac's delta function.
The cumulative distribution of areas is given as $\Theta_i(s) =
\int_{-\infty}^{s} h_i(w)dw$.  In order to obtain an interpolated and
smoothed version of $\Theta_i(s)$, this function is convolved with a
regularizing Gaussian kernel yielding $B_i(s)= \Theta_i(s)* g_\sigma
(s)$, where $g_\sigma$ is the normal distribution with standard
deviation $\sigma$ and $*$ stands for the convolution operation.  Now,
by making $H_i(s)=log(B_i(s))$, the fractal dimension of node $i$ as
an adimensional function of the spatial scale is given by
Equation~\ref{eq:frac}.

\begin{equation} 
f_i(s)=d \, H_i(s)/ds =B'_i(s)/B_i(s)  \label{eq:frac}
\end{equation}

\begin{figure}
  \begin{center} 
  \includegraphics[angle=-90,scale=0.25]{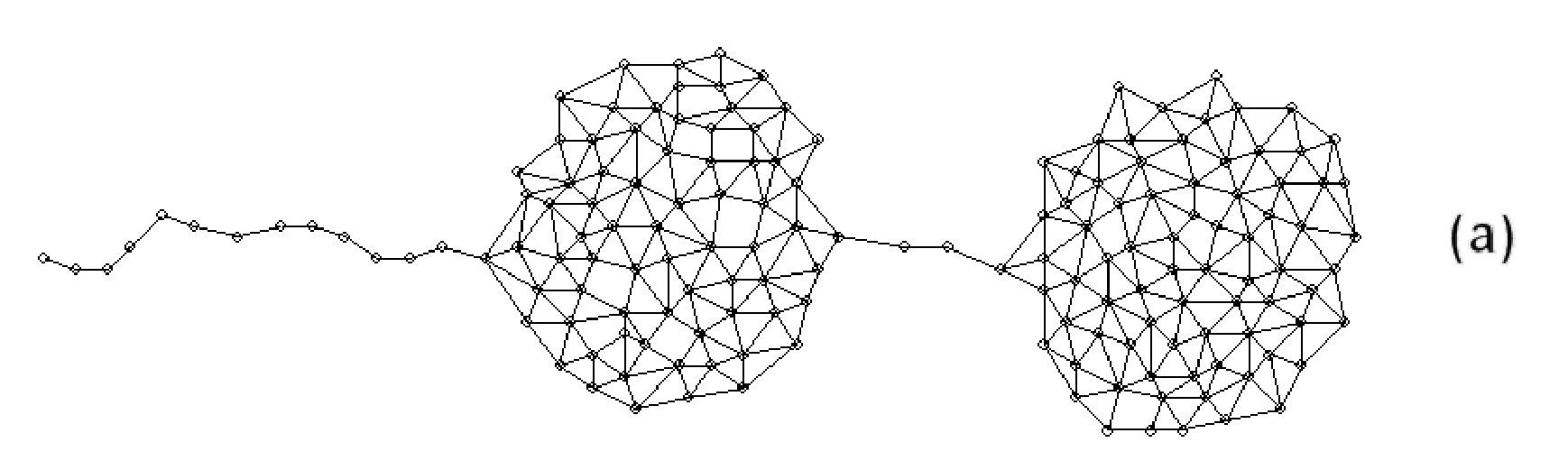}
  \includegraphics[angle=-90,scale=1]{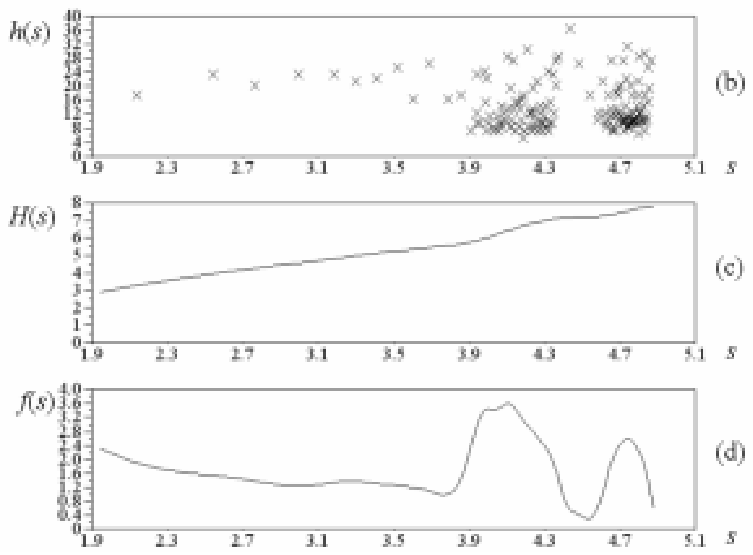}
  \end{center} 
  \caption{A Voronoi network (a) and its respective area distribution
  (b), logarithm of cumulative areas (c) and fractal dimension in
  terms of the spatial scale $s$ (d).~\label{fig:ex_frac}}   
\end{figure}

The higher the value of $f_i(s)$, the faster the accumulated area
increases along the subsequent shortest paths at the respective
spatial scale $s$.  Given that the areas of the Voronoi cells are
added at once whenever their respective distances are reached, fractal
dimensions larger than 2 can be observed.  Similarly, the adoption of
edge costs smaller than the respective Euclidean distance also leads
to `\emph{superfractal}' dimensions, in the sense that its dynamics
while it undergoes inundation exceeds the topological dimension of the
space where the network is embedded.  Observe that it is also possible
to obtain fractal dimensions starting from a \emph{set} of points
instead of a single point.  Figure~\ref{fig:ex_frac} shows a node
spatial distribution (a), the area distribution obtained from the
respective Voronoi tessellation (b), the cumulative distribution (c)
and the fractal dimension (d), all in terms of the spatial scale $s$.
Such results were obtained by starting the network inundation from the
leftmost point in (a). While each local minimum of $f_i (s)$
identifies a bottleneck along the network topology, as clearly
illustrated in this example, the average fractal value $\left< f
\right>$ taken along all possible scales provides an interesting
quantification of the overall spatial connectivity of the nodes in the
net.

\begin{figure}
  \begin{center}
  \includegraphics[angle=-90,scale=0.45]{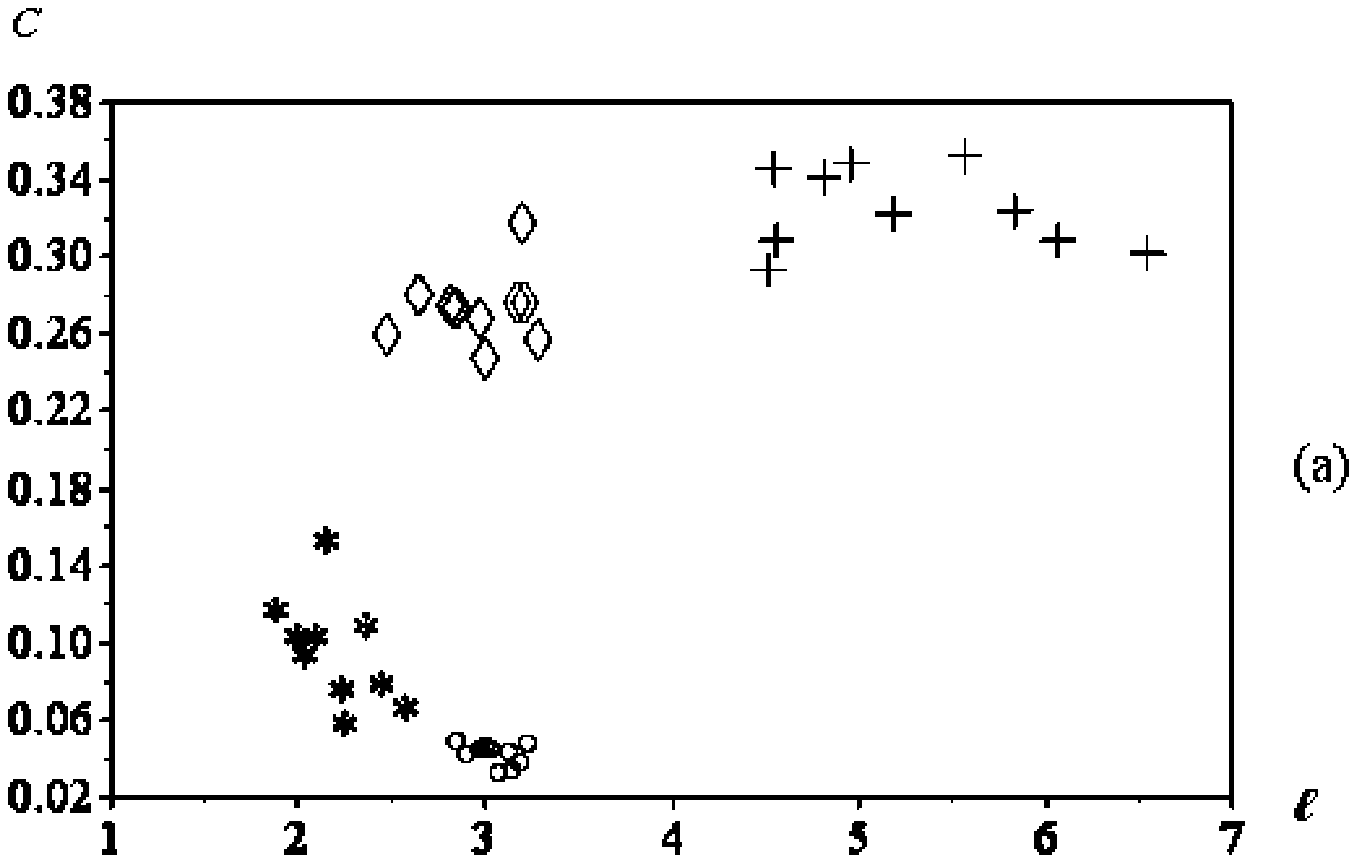}
  \includegraphics[angle=-90,scale=0.45]{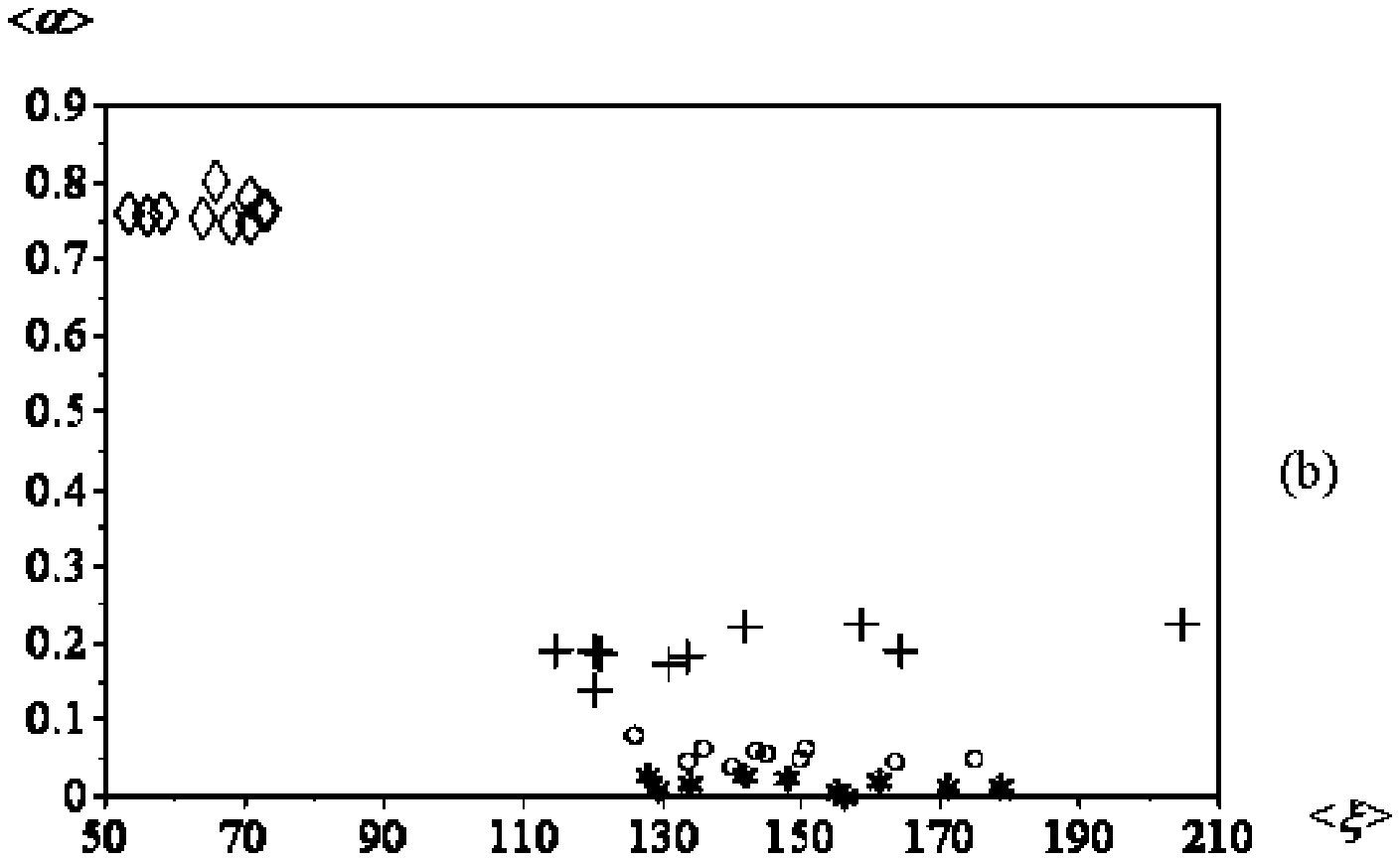}
  \includegraphics[angle=-90,scale=0.45]{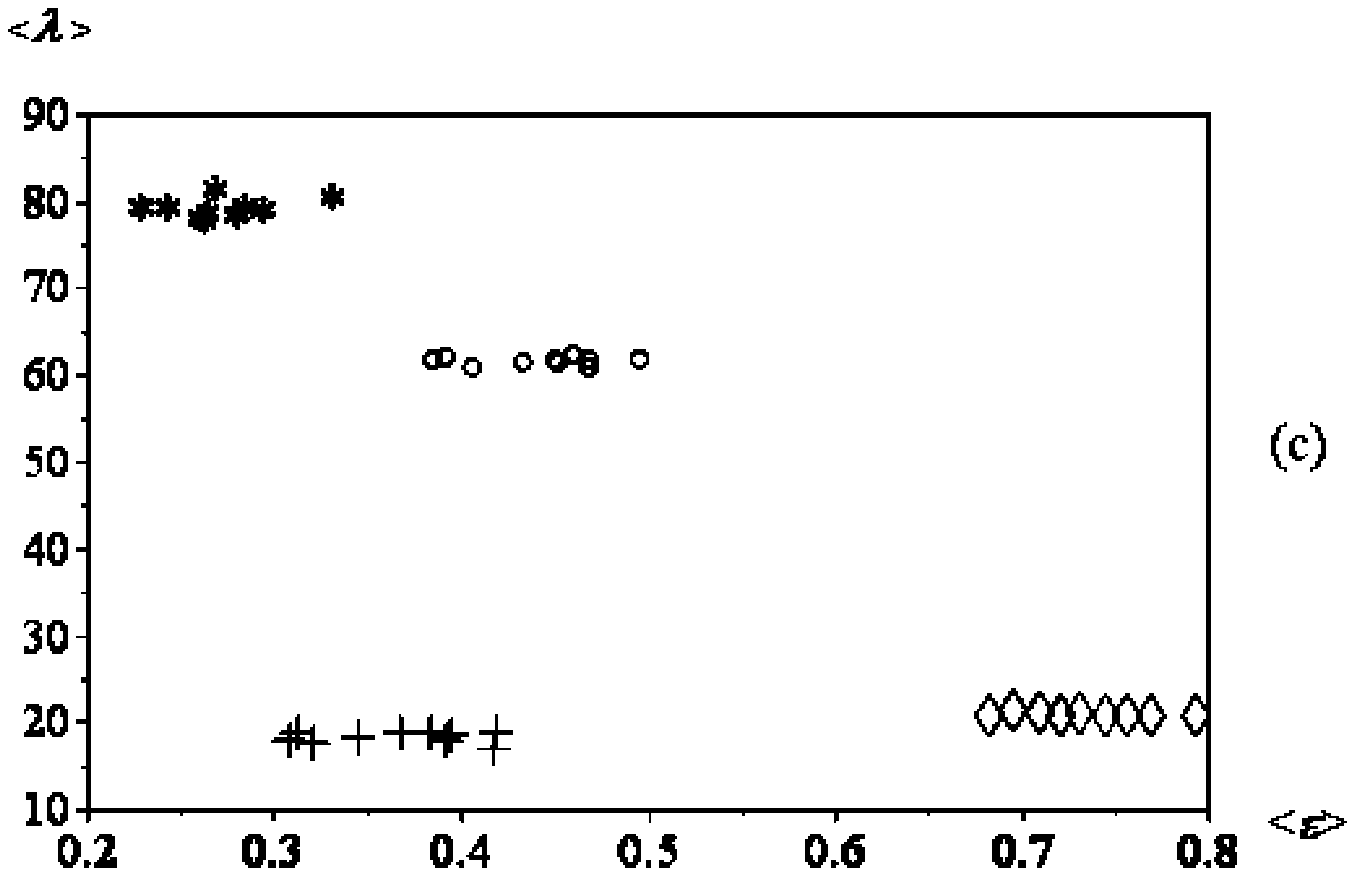}
  \includegraphics[angle=-90,scale=0.45]{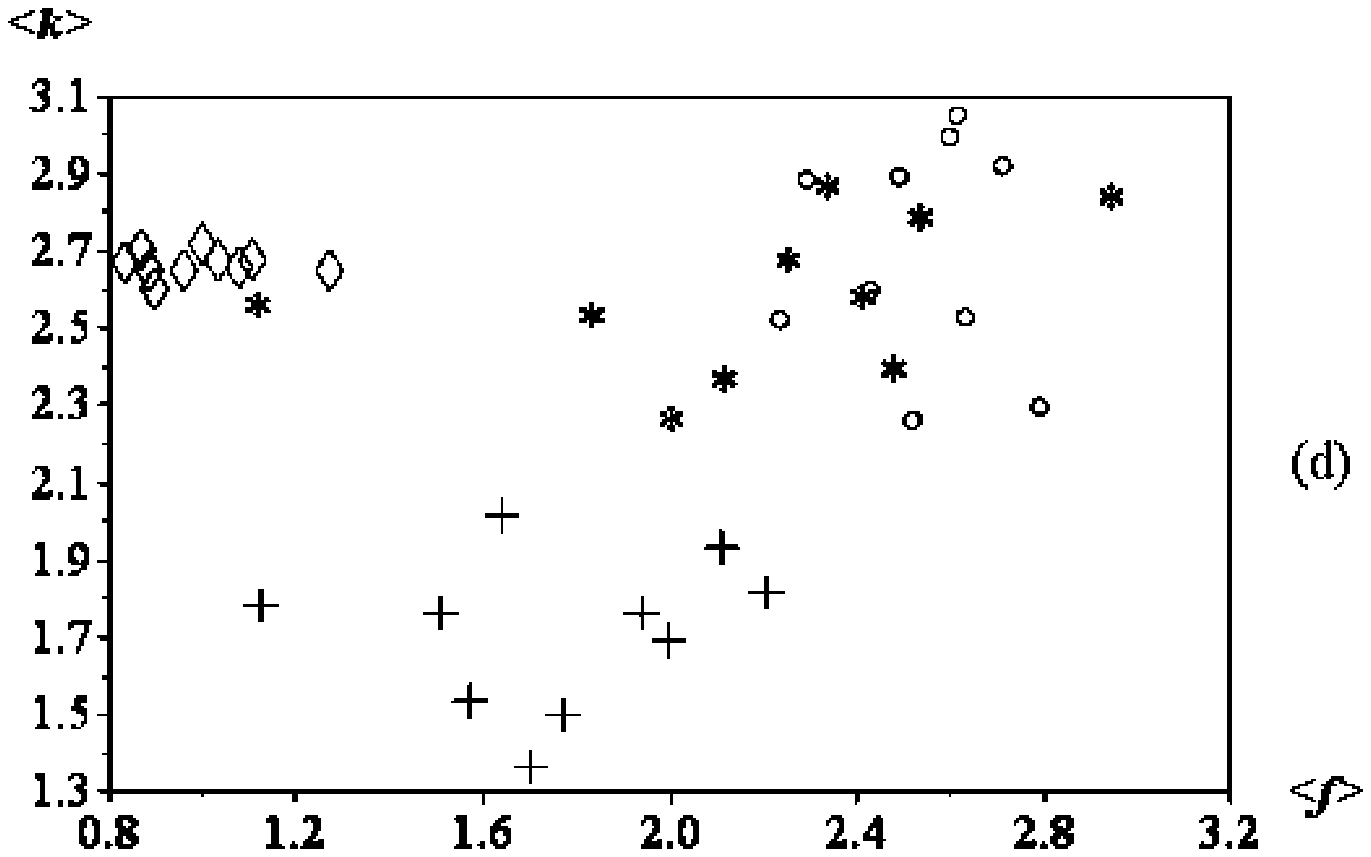} 
  \end{center}
  \caption{Phase space distributions of pairs of average measurements
  obtained from several realizations of the considered network models
  (random = $\circ$, Watts-Strogatz = $\diamond$, Albert-Barab\'asi =
  $\star$ and fractal random = $+$).~\label{fig:phases}} 
\end{figure}

The four diagrams in Figure~\ref{fig:phases} show the distribution of
the above suggested measurements, taken pairwise, experimentally
obtained from several realizations of each network model considering
$L=121$ and $\gamma=0.02$.  For comparison's sake, all simulations had
number of edges as close as possible to $n=350$, as well as number of
nodes $N \cong 140$.  In the case of the fractal random network, we
used $M=5$, $p_0=0.00035$ and $\gamma=0.5$.  For Watts-Strogatz,
$p=10\gamma$, with the initial configuration obtained by directly
connecting all spatially adjacent nodes.  The Albert-Barab\'asi model
used $m_0=10$ and $m=3$, and $M=5$ and $p_0=0.000034$ were adopted for
the fractal random networks.  Distinctively segregated clusters were
obtained in most cases, except for (d), and the fractal random
measurements resulted generally more dispersed than those obtained for
the other models.  This is a consequence of the relatively low number
of nodes $M=5$ required in order that the total number of network
nodes results similar to those of the other simulated models.

From figure~\ref{fig:phases}(a) we have that the fractal random
network led to the highest overall cluster coefficients $C$ and
average lengths $\ell$ values.  The adjacencies in (b) increase along
the following sequence: Albert-Barab\'asi, random, fractal random and
Watts-Strogatz.  The distinctively high adjacency values and low
shortest distance values exhibited by the Watts-Strogatz model is a
direct consequence of the initial adjacent configuration used in that
case.  The more intense spatial regularity of that model also led to
high effective length values and low edge distances in the phase space
in (c), which is also marked by low dispersion of the average edge
distances.  The fractal random network yielded effective distances
slightly higher than those obtained for the Albert-Barab\'asi network,
which was also characterized by small edge distances, which reflects
the larger number of small spatial scale Voronoi cells.  Indeed, the
spatially adjacent and spatially uniformly distributed connections in
the Watts-Strogatz case also implied most measurements to be less
dispersed.  As shown in (d), the random and Albert-Barab\'asi networks
behave similarly with respect to average fractal dimension and node
degree, whose values are the highest among the considered network
models.  The fractal random network led to the smallest node degrees
and intermediate fractal dimensions.  The spatially more localized
connections in the Watts-Strogatz network, makes its inundation
smoother, implying the lowest fractal dimension values.  Except for
this network, a positive correlation is observed in (d), reflecting
the fact that higher node degrees tends to favor higher fractal
dimensions.  Another interesting point in (d) is that the less
spatially localized connections in the random and Albert-Barab\'asi
networks induced superfractal behavior.  As a matter of fact, the
relatively smaller average fractal dimensions presented by the
Watts-Strogatz and fractal random models can be taken as an indication
that they are more intensely influenced by spatial constraints, as is
indeed the case.

The above results illustrate the potential of the suggested network
measurements for quantifying and analysing the properties of networks
as far as the possible influence of spatial constraints are concerned,
providing a comprehensive integrated framework for investigating real
complex networks.  Of particular interest are the local/global
hierarchical and topographical organization of mammals' cortical
structures, the relationship between neuronal function and
connectivity and shape \cite{Costa_BM:2003}, as well as the
spatio-temporal gene activation patterns underlying animal
development.  Other interesting possibilities are to consider fractal
networks where the subsequent partition of the Voronoi is done with
probability reflecting the degree of the parent node and/or additional
spatial constraints such as varying density along $\Omega$.

\begin{acknowledgments}
The author is grateful to FAPESP (processes 99/12765-2 and 96/05497-3)
and Human Frontier Science Program for financial support.
\end{acknowledgments}

\bibliography{spatnet}

\end{document}